\documentclass{article}
\usepackage{amssymb,amsbsy,amsmath,amsfonts,amssymb,amscd}
\usepackage{graphicx,epsfig,times,epsf,subfigure}

\newcommand{\Z}{\mathbb{Z}}

\newcommand{\R}{\mathbb{R}}
\newcommand{\C}{\mathbb{C}}

\newcommand{\brE}{\mathbf{E}}

\newcommand{\brj}{\mathbf{j}}
\newcommand{\brk}{\mathbf{k}}

\newcommand{\brs}{\mathbf{s}}

\newcommand{\eps}{\varepsilon}

\newcommand{\RE}{\Re\it{e}}

\newcommand{\dfl}{\lfloor}
\newcommand{\ffl}{\rfloor}

\begin{document}

\title{Optical scattering by a nonlinear medium, II: induced photonic crystal in a nonlinear slab of BBO}

\author{Pierre Godard, Fr\'ed\'eric Zolla, Andr\'e Nicolet \vspace{.2cm} \\
\small{Institut Fresnel UMR CNRS 6133, Facult\'e de Saint-J\'er\^ome case 162} \\
\small{13397 Marseille Cedex 20, France} \\
\small{pierre.godard@fresnel.fr,frederic.zolla@fresnel.fr,andre.nicolet@fresnel.fr}}

\date{}


\maketitle
\begin{abstract}
The purpose of this paper is to investigate the scattering by a
nonlinear crystal whose depth is about the wavelength of the
impinging field. More precisely, an infinite nonlinear slab is
illuminated by an incident field which is the sum of three plane
waves of the same frequency, but with different propagation vectors
and amplitudes, in such a way that the resulting incident field is
periodic. Moreover, the height of the slab is of the same order of
the wavelength, and therefore the so-called slowly varying envelope
approximation cannot be used. In our approach we take into account
some retroactions of the scattered fields between them (for
instance, we do not use the nondepletion of the pump beam). As a
result, a system of coupled nonlinear partial differential equations
has to be solved. To do this, the finite element method (FEM)
associated with perfectly matched layers is well suited.
Nevertheless, when using the FEM, the sources have to be located in
the meshed area, which is of course impossible when dealing with
plane waves. To get round this difficulty, the real incident field
is simulated by a virtual field emitted by an appropriate antenna
located in the meshed domain and lying above the obstacle (here the
slab).
\end{abstract}

\section{Introduction}

The development of photonic science in nanotechnologies requires an
always increasing control of light. Surface-phenomena, metamaterials
or the use of nonlinear optics are very efficient ways to do this.
In this paper, we combine the last two options: an electromagnetic
field induces, in a homogeneous nonlinear medium, a periodicity of
period close to the considered wavelength. A precise description of
the system is given in the next section.

The majority of work in nonlinear optics applies to the
\emph{propagation} of a wave in a nonlinear medium (see
\cite{aKivshar00} for a review
), that is, the wavelength is small compared to the length of the
path of light in the nonlinear medium. In this case, the paraxial
approximation is often used and the equations obtained are of
parabolic kinds. For example, when studying the propagation of a
soliton in an optical fiber (say, oriented along the $z$-axis), one
usually restrict the problem in the $(z,t)$-plane, neglecting the
transversal effects. This leads to equations of the nonlinear
Schr\"{o}dinger type
. Some works have been done to determine the limits of this approach
(\cite{aChen97,aCiattoni05,aDrouart08}). This justifies the studies
done directly with Maxwell's equations, as in
\cite{aFerrando03,aDarmanyan01}, where transversal effects are taken
into account.


On the other side, when the wavelength is far larger than the
obstacle, a mean-field approximation is used (\cite{Cen,aXie03}).
The purpose of this paper is to stand between these two states, in
the realm of resonance, where rough approximations cannot be
applied. To this end, we use the FEM: the precision in the change of
apparent permittivity inside the slab (i.e., the inhomogeneity of
the sources, resonating with different frequencies, of the total
field) is only dictated by the memory size of the computer.

\vspace{.2cm}

The study of nonlinear optics in periodic material is of course not
new (\cite{aBloembergen70,aReinisch83}, see also \cite{bNeviere00}).
It is known that Kerr photonic crystals allow to produce systems
with hysteresis (\cite{Cen,aSoljacic02}), or with a transmission
range that depends on the amplitude of the fields
(\cite{aCicek08,aFujisawa04}). Peak of transmission can also appear
in the gap due to the nonlinearity (\cite{Chen,John}).

When considering harmonic generation due to a nonlinear periodic
structure, the phenomena that appear in photonic crystals are still
richer, for the frequency components of the fields can have
completely different behaviors
(\cite{aLifshitz04,aCenteno06,aKlein07,aMattiucci07}).

\vspace{.2cm}

Finally, this paper also aims at exposing an application of the
companion paper \cite{aGodard11a}, that gives a new route to obtain
propagation equations in nonlinear optics.

\section{Set up of the problem}

\subsection{Description of the system}

\begin{figure}
\centering
  \includegraphics[width=.8\textwidth]{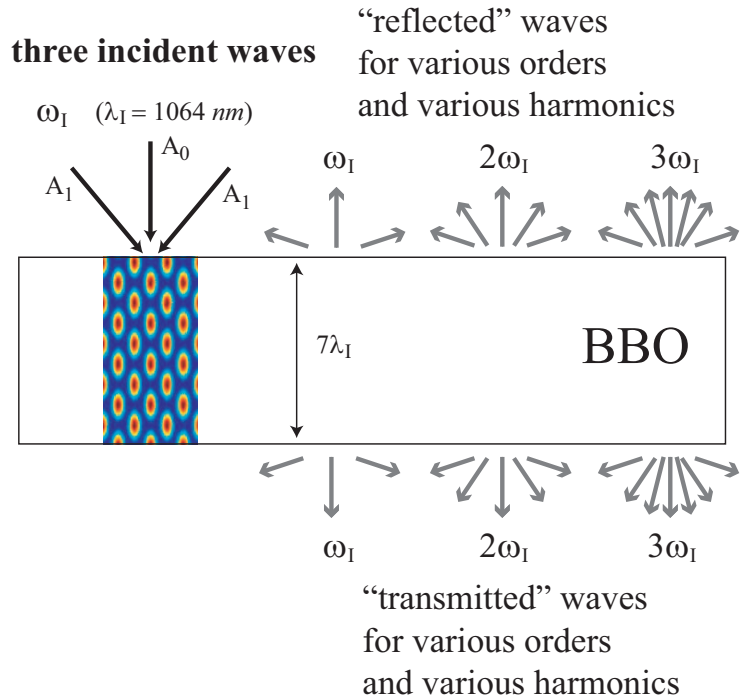}
\caption{The geometry of the system is a slab, invariant along the
horizontal direction, called the $x$ axis, and along the direction
perpendicular to the plane of the sheet of paper, called the $z$
axis. Made up of $\beta$ Barium Borate (BBO), it is illuminated by
three plane waves: one with a normal incidence and an amplitude
$A_0$, and two with a symmetric oblique incidence and amplitude
$A_1$. The interferences between these different incident waves
oscillating at $\omega_I$, induces a periodic structure, equivalent
to a diffraction grating or a finite photonic crystal. This is seen
in color inside the slab, where the pattern of the intensity field
$\|\brE_1^i\|$ of the incident waves is shown. This nonlinear
induced grating generate some harmonics. The number of propagating
order increases with the pulsation. (Three periods are presented in
this picture, though the simulation runs with one period only.)}
  \label{fig_grating_geom}
\end{figure}

The choice of the test structure has been dictated by two
guidelines: it must be at the same time simple enough to lead to
tractable numerical models and possibly feasible experiments, and
enhance the nonlinear effects both qualitatively and quantitatively.
For this, we propose the following experiment: let three plane waves
impinge on a slab, made up of a nonlinear and non-centro-symmetric
crystal. One wave is directed normally with respect to the slab, the
other two are symmetrically oriented with respect to it and have the
same amplitude (the analytical expression of the incident field is
given in subsection \ref{SubsecPractPointView}). In this way, the
problem is periodic, along one direction that we call the $x$-axis.
This incident field create an optical lattice
(\cite{aFreedman06,aZhang09}), as seen in the figure
\ref{fig_grating_geom}. Hence, the scattered field presents several
orders of reflection and of transmission. Furthermore, the scattered
field oscillate at several pulsations, due to harmonic generation in
the crystal. The range of ratios \emph{wavelength/period of the
crystal} offers a variety of response for the different harmonics.
In particular, the smaller the wavelength, the more order excited.
This implies that there are more scattered angles for the fields
oscillating at the generated harmonics than for the one oscillating
at the incident frequency. Moreover, and we find it spectacular, the
energy that flows along each direction of the crystal is not at all
a monotonous function of the incident intensity. This offers new
ways to control the directions along which the higher harmonics
escape from the nonlinear medium.

Let $\lambda_I$ be the wavelength of the incident field. The slab we
choose had a thickness of $7.06\ \lambda_I$. In this range, the
slowly varying envelop is not a good approximation, so we have to
solve the complete set of Maxwell's equations. But, as shown in
\cite{aGodard11a}, Maxwell's equations lead to an infinite set of
nonlinear and coupled equations. The article just mentioned is
mainly devoted to the elaboration of this system, and then its
truncation, necessary for a numerical study. Some results are
briefly reminded in the following subsection; the reader is reported
to \cite{aGodard11a} for their justification.

\subsection{What kind of equations has to be solved?}

We consider a spatially local, nonbianisotropic, stationary,
magnetically linear, smooth medium (see \cite{aGodard11a} for a
definition; the idea behind a smooth medium is that we can make a
Taylor expansion of the electric induction field in function of the
electric field). For simplicity, the electric susceptibility tensor
$\chi^{(n)}$ are neglected when $n>3$. The nonlinear effects we want
to consider are mainly the second and third harmonic generations.
Hence the nonharmonic processes (like Raman or Brillioun
scatterings) are not taken into account.

We write the electric vector, at a point $\brs$ and a time $t$ as

$$\brE(\brs,t)=2\RE\{\brE_1(\brs)e^{-i\omega_It}+\brE_2(\brs)e^{-2i\omega_It}+\brE_3(\brs)e^{-3i\omega_It}\}.$$
Note that this expression implies that we neglect the cascading
effect that create harmonics higher than the third one; also, as it
is often the case, the static component is supposed to vanish.

To write the propagation equations, we need to introduce two
notations: the first one is the operator $\mathcal{M}_p^{lin}$,
that, when applied to the $p$-th component $\brE_p$ of the electric
field, gives the equation satisfied by $\brE_p$ in a linear medium;
the second one is a short way of writing the interactions between
several components of the total electric field. For example,
$\dfl\brE_1,\brE_1\ffl:=\chi^{(2)}(\omega_I,\omega_I)\brE_1\brE_1$,
describes the interaction of $\brE_1$ with itself - this term is the
source of the second harmonic generation -,
$\dfl\brE_{-1},\brE_1,\brE_1\ffl:=\chi^{(3)}(-\omega_I,\omega_I,\omega_I)\brE_{-1}\brE_1\brE_1$,
is the term that describes the interaction between $\brE_{-1}$,
$\brE_1$ and $\brE_1$. Since $\brE_{-p}=\overline{\brE_p}$, this is
precisely the optical Kerr effect. More formally, we have

$$\mathcal{M}_p^{lin}(\brE_p):=-\eps_0^{-1}\nabla\times(\mu^{-1}(p\omega_I)\nabla\times
\brE_p)+(p\omega_I)^2\eps_r^{(1)}(p\omega_I)\brE_p$$%
and

$$\dfl\brE_{p_1},\cdots,\brE_{p_n}\ffl:=\chi^{(n)}(p_1\omega_I,\cdots,p_n\omega_I)\brE_{p_1}\cdots\brE_{p_n}$$
where $p_i\in\Z$. Note that this term contributes to the component
of the $n$-th order polarization oscillating at the pulsation
$(p_1+\cdots+p_n)\omega_I$

Now, as said above, we want to study the second and third harmonic
generations. Their respective sources are $\dfl\brE_1,\brE_1\ffl$
for $\brE_2$, and $\dfl\brE_1,\brE_2\ffl$ and
$\dfl\brE_1,\brE_1,\brE_1\ffl$ for $\brE_3$. Note that we consider
here a cascading effect: $\brE_1$ and $\brE_2$ interact to create
$\brE_3$. In fact, we give a set of equations that allows to
consider a lossless process (\cite{aGodard11a,bGodard10}):

\begin{subequations}\label{SystemEqOrdre3Degre3E0nul}
\begin{align}
\mathcal{M}_1^{lin}&(\brE_1)\label{SystemEqOrdre3Degre3E0nulE1} \\
&+\omega_I^2\Big(2\dfl\brE_{-2},\brE_3\ffl+2\dfl\brE_{-1},\brE_2\ffl\nonumber \\
&+6\dfl\brE_{-3},\brE_1,\brE_3\ffl+3\dfl\brE_{-1},\brE_{-1},\brE_3\ffl+3\dfl\brE_{-3},\brE_2,\brE_2\ffl\nonumber \\
&+6\dfl\brE_{-2},\brE_1,\brE_2\ffl+3\dfl\brE_{-1},\brE_1,\brE_1\ffl\Big)=\mathfrak{s}(\brE^i_1),\nonumber \\
\nonumber \\
\mathcal{M}_2^{lin}&(\brE_2)\label{SystemEqOrdre3Degre3E0nulE2} \\
&+(2\omega_I)^2\Big(2\dfl\brE_{-1},\brE_3\ffl+\dfl\brE_1,\brE_1\ffl\nonumber \\
&+6\dfl\brE_{-3},\brE_2,\brE_3\ffl+6\dfl\brE_{-2},\brE_1,\brE_3\ffl+3\dfl\brE_{-2},\brE_2,\brE_2\ffl\nonumber \\
&+6\dfl\brE_{-1},\brE_1,\brE_1\ffl\Big)=0,\nonumber \\
\nonumber \\
\mathcal{M}_3^{lin}&(\brE_3)\label{SystemEqOrdre3Degre3E0nulE3} \\
&+(3\omega_I)^2\Big(2\dfl\brE_1,\brE_2\ffl\nonumber \\
&+3\dfl\brE_{-3},\brE_3,\brE_3\ffl+6\dfl\brE_{-2},\brE_2,\brE_3\ffl+6\dfl\brE_{-1},\brE_1,\brE_3\ffl\nonumber \\
&+3\dfl\brE_{-1},\brE_2,\brE_2\ffl+\dfl\brE_1,\brE_1,\brE_1\ffl\Big)=0.\nonumber
\end{align}
\end{subequations}

Note that some other interesting effects are taken into account: the
optical Kerr effect ($\dfl\brE_1,\brE_1,\brE_{-1}\ffl$), the
depletion of the pump beam ($\dfl\brE_2,\brE_{-1}\ffl$). On the
other hand, some terms are really weak (typically
$\dfl\brE_3,\brE_3,\brE_{-3}\ffl$ as compared to
$\dfl\brE_1,\brE_1,\brE_1\ffl$) and do not have important
consequences on the fields. The presence of these interactions is
only due to the energy conservation property, which is a good test
from the numerical point of view. Finally, $\mathfrak{s}(\brE^i_1)$
is the external source of $\brE_1$; since in our case the slab is
illuminated by plane waves, $\mathfrak{s}(\brE^i_1)$ vanishes.

\section{Some Numerical Results}

\subsection{From a practical point of
view}\label{SubsecPractPointView}

To solve the system of equations we posed, numerical methods have to
be used. The nonlinearity is tackled by a Newton-Raphson scheme and,
at each step, the finite element method was chosen for its ability
to treat the nonhomogeneous sources, induced by the nonlinearity, of
each equation. The incident wave is monochromatic,
$\brE^i(\brs,t)=2\RE\{\brE_1^i(\brs)e^{-i\omega_It}\}$, and to
simplify the comparison with an experimental setup, we fix
$\omega_I$ such that the associated wavelength is
$\lambda_I=1064\,nm$, which corresponds to a Nd:YAG laser.

Only two-dimensional problems are addressed here. We recall that the
physical system is invariant along the $x$ and the $z$ axis. The
$TH$ polarization is chosen, that is, a function
$u^i:\R^2\rightarrow\C$ satisfies, in Cartesian coordinates,

$$\brE^i_1(x,y,z)=u^i(x,y)\hat{z}.$$
Now a key step is described: the crystal we choose is the $\beta$
Barium Borate (from now on denoted as BBO); it is known that there
exists one orientation of the crystal that guarantees that the total
field has the same polarization as the incident field:

$$\brE_p(x,y,z)=u_p(x,y)\hat{z}$$
for some functions $u_p:\R^2\rightarrow\C$. We choose this
orientation, so that the $p$-th component of the electric field is
described with these scalar functions.

Finally, the propagation equation system
(\ref{SystemEqOrdre3Degre3E0nul}) being given in term of the total
electric field, we use a virtual antenna to simulate the incident
field (\cite{aGodard08}). Perfectly matched layers are used to
impose outgoing wave conditions above and below the
slab\footnote{From the physical point of view, only the scattered
part of $u_1$ satisfies an outgoing wave condition. But the
principal of the virtual antenna is precisely to find a source,
located in the meshed domain, that generates, in the neighborhood of
the obstacle, exactly the incident field. All the details are given
in \cite{aGodard08}.}, for $u_1$, $u_2$ and $u_3$. For the reader's
convenience, we give the electric current that simulates the
incident electric field:

$$\brj(x,y,t)=2\RE\{\j(x,y)\ e^{-i\omega_I t}\}\hat{z},$$
with

$$\j(x,y)=\frac{1}{\mu\ \omega_I}\{\ k A_0e^{-iky}+2\beta A_1\ e^{-i\beta y}\cos(\alpha
x)\}\delta(y-h),$$ to be evaluated at $h$, where $h$ is the height
of the thread on which the current flows,
$\alpha:=\brk\cdot\hat{x}=k \sin(\theta)$ and
$\beta:=\brk\cdot\hat{y}=k \cos(\theta)$
($k:=\|\brk\|=2\pi/\lambda_I$ is common to the three incident waves
and $\theta$ is the obliquity angle of the waves with the amplitude
$A_1$), and $\delta$ is the Dirac distribution. The expression of
the incident field is then

$$u^i(x,y)=A_0e^{-iky}+2A_1e^{-i\beta y}\cos(\alpha x).$$

\subsection{Scattering Of Three Waves On A Slab}

We come now to the results of the simulation of the experiment
described in the figure \ref{fig_grating_geom}. 
This system exhibits a non
trivial and quite complex behavior because of the induced
diffraction grating.

The numerical value of the relevant susceptibility components are
given in the table \ref{tabXiBBO}.
We note that they do not depend on the frequency.

\begin{table}[!]
\begin{center}
\begin{tabular}{|l|l|l|l|l|}
  \hline
  $\lambda$ & $n_e$ & $\chi^{(1)\,z}_{\qquad z}$ & $\chi^{(2)\,z}_{\qquad zz}$ & $\chi^{(2)\,z}_{\qquad zzz}$ \\
  \hline
  $1064$ & $1.5426$ & $1.3793$ & $2.10^{-12}$ & $10^{-23}$ \\
  $532$ & $1.5555$ & $1.4196$ & $2.10^{-12}$ & $10^{-23}$ \\
  $355$ & $1.5775$ & $1.4885$ & $2.10^{-12}$ & $10^{-23}$ \\
  \hline
\end{tabular}
\end{center}
\caption{The wavelength, in $nm$, the refraction index of the
extraordinary axis, and the values of the relevant (in that article)
$\chi^{(1)}$, $\chi^{(2)}$ and $\chi^{(3)}$, respectively without
unit, in $m.V^{-1}$ and in $m^2.V^{-2}$.} \label{tabXiBBO}
\end{table}

We are interested in the directions where the fields scatter. Out of
the slab, for each frequency, the electric field satisfies a
Helmholtz equation. We write its propagating solution as

\begin{align*}
\brE_p(x,y)&=\sum_{n\in\mathcal{U}_p}\{b^{(r)}_{p,n}e^{i\big(\frac{2\pi
n}{d}x+(k_p^2-(\frac{2\pi n}{d})^2)^{1/2}y\big)} \\
&\quad+b^{(t)}_{p,n}e^{i\big(\frac{2\pi n}{d}x-(k_p^2-(\frac{2\pi
n}{d})^2)^{1/2}y\big)}\}\hat{z},
\end{align*}
where $b^{(r)}_{p,n}$ (resp. $b^{(t)}_{p,n}$) denotes the
coefficient of the $n$-th order of the reflected (resp. transmitted)
wave at pulsation $p\omega_I$, and
$\mathcal{U}_p:=\{n\in\Z:k_p^2-(\frac{2\pi n}{d})^2>0\}$. Since
$k_p=p\omega_I/c$ (where $c$ is the light velocity), the higher the
harmonic $p$, the higher the modulus of the wave vector $k_p$; in
other words, the size of $\mathcal{U}_p$ increases with $p$. This
means that there are more scattered angles for higher harmonics than
for the field oscillating at the fundamental frequency. In our
particular case, whereas only the orders $-1$, $0$ and $1$ are
present in $\brE_1$, the third harmonic contains orders from $-3$ to
$3$. These ranges depend on the wavelength and the inclination
$\theta$ of the incident beams; had we chosen a larger $d$, the
propagating orders allowed for the first harmonic would have ranged
from, say, $-2$ to $2$.

\begin{figure}
\centering
          \subfigure[$e^{(t)}_{1,0}$]
             {\includegraphics[width=0.40\textwidth]{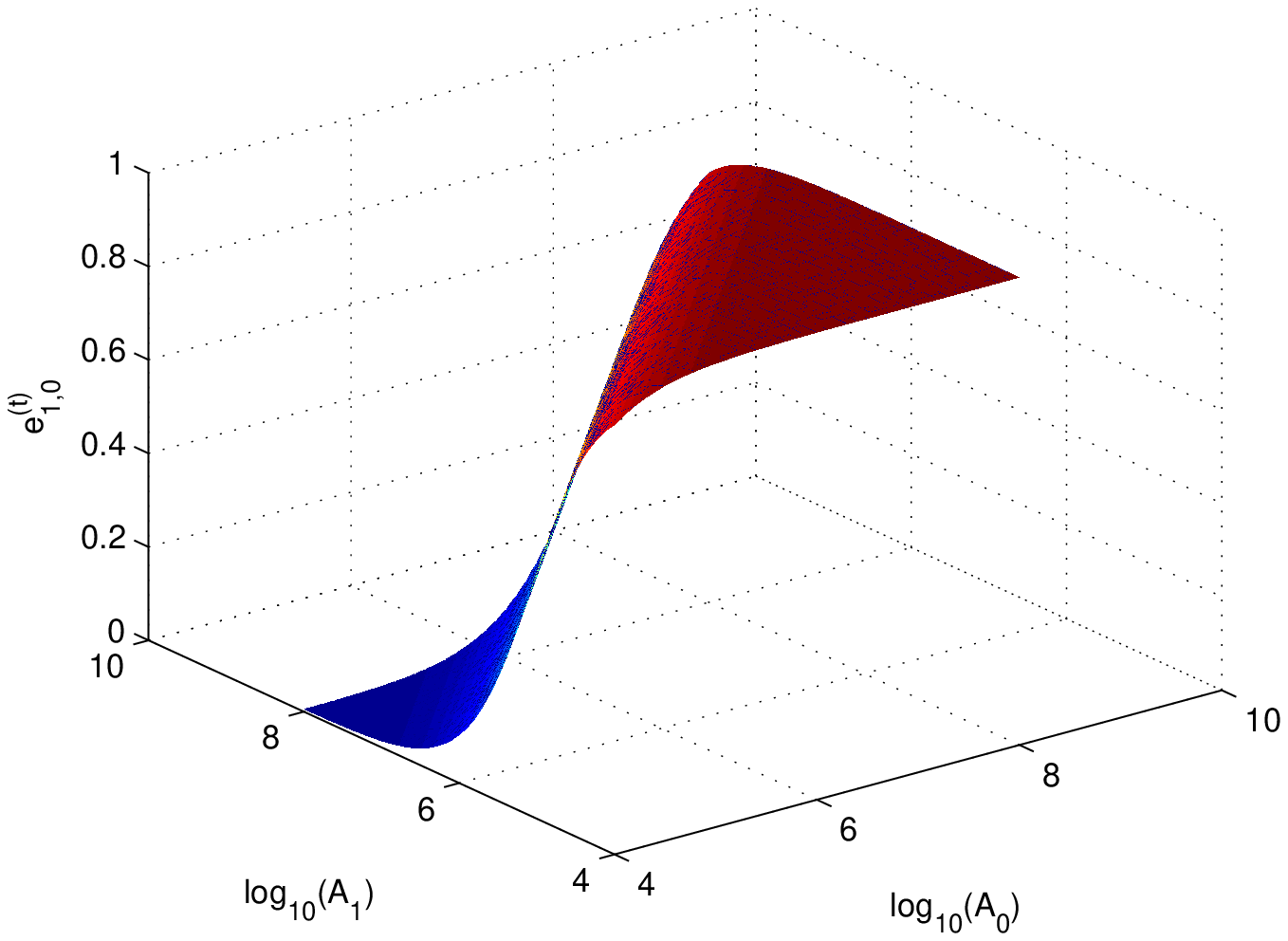}}
          \subfigure[$e^{(t)}_{1,1}$]
             {\includegraphics[width=0.40\textwidth]{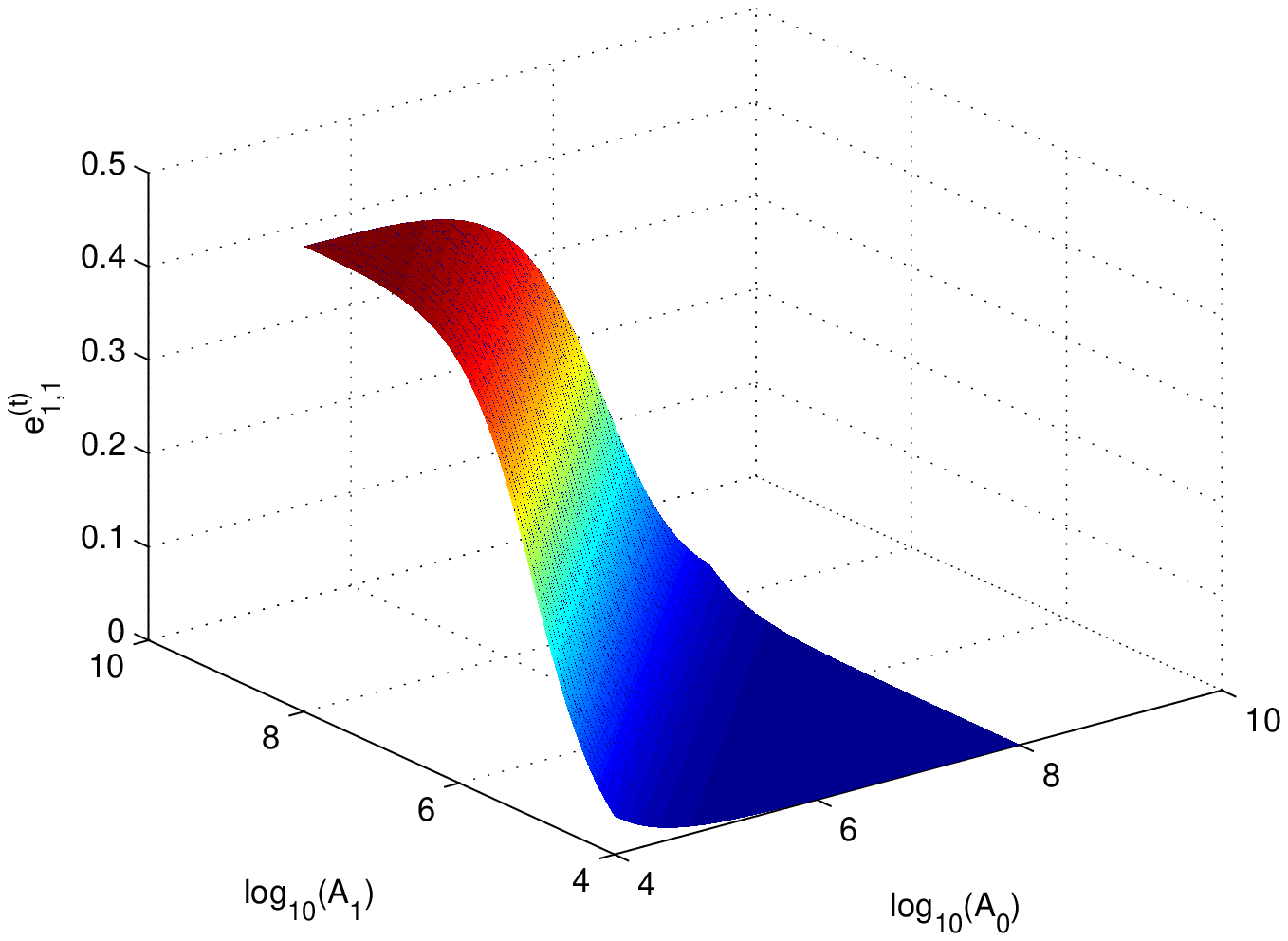}}
       \caption{The efficiency of the transmitted propagating waves at the frequency $\omega_I$. As is easily understood, when $A_1$ tends to zero, all the light escape perpendicularly to the slab, i.e., $e^{(t)}_{1,0}$ tends to one and $e^{(t)}_{1,1}$ vanishes.}
       \label{figSlab1t}
\end{figure}

\begin{figure}
\centering
          \subfigure[$e^{(t)}_{2,0}$]
             {\includegraphics[width=0.40\textwidth]{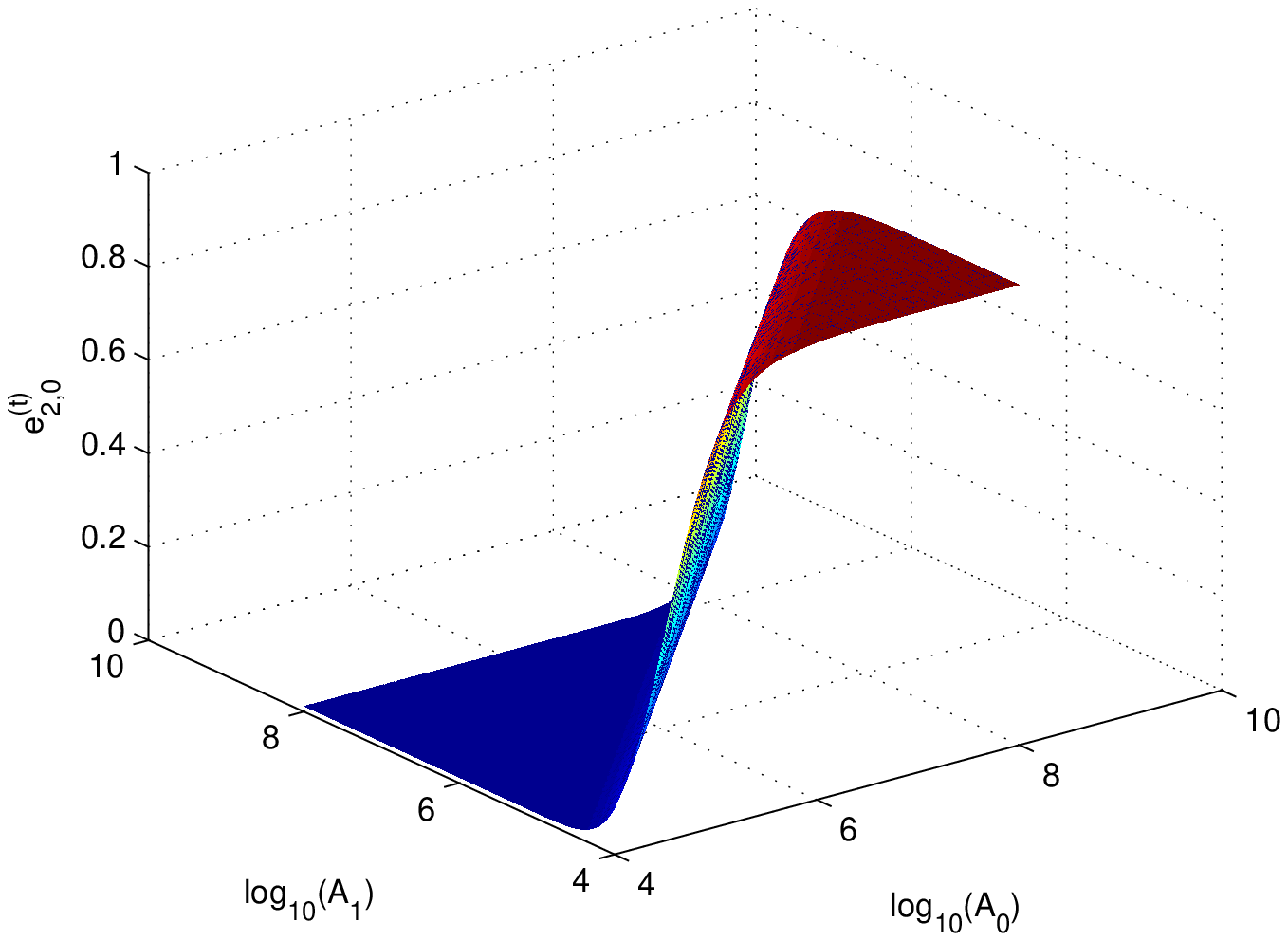}}
          \subfigure[$e^{(t)}_{2,1}$]
             {\includegraphics[width=0.40\textwidth]{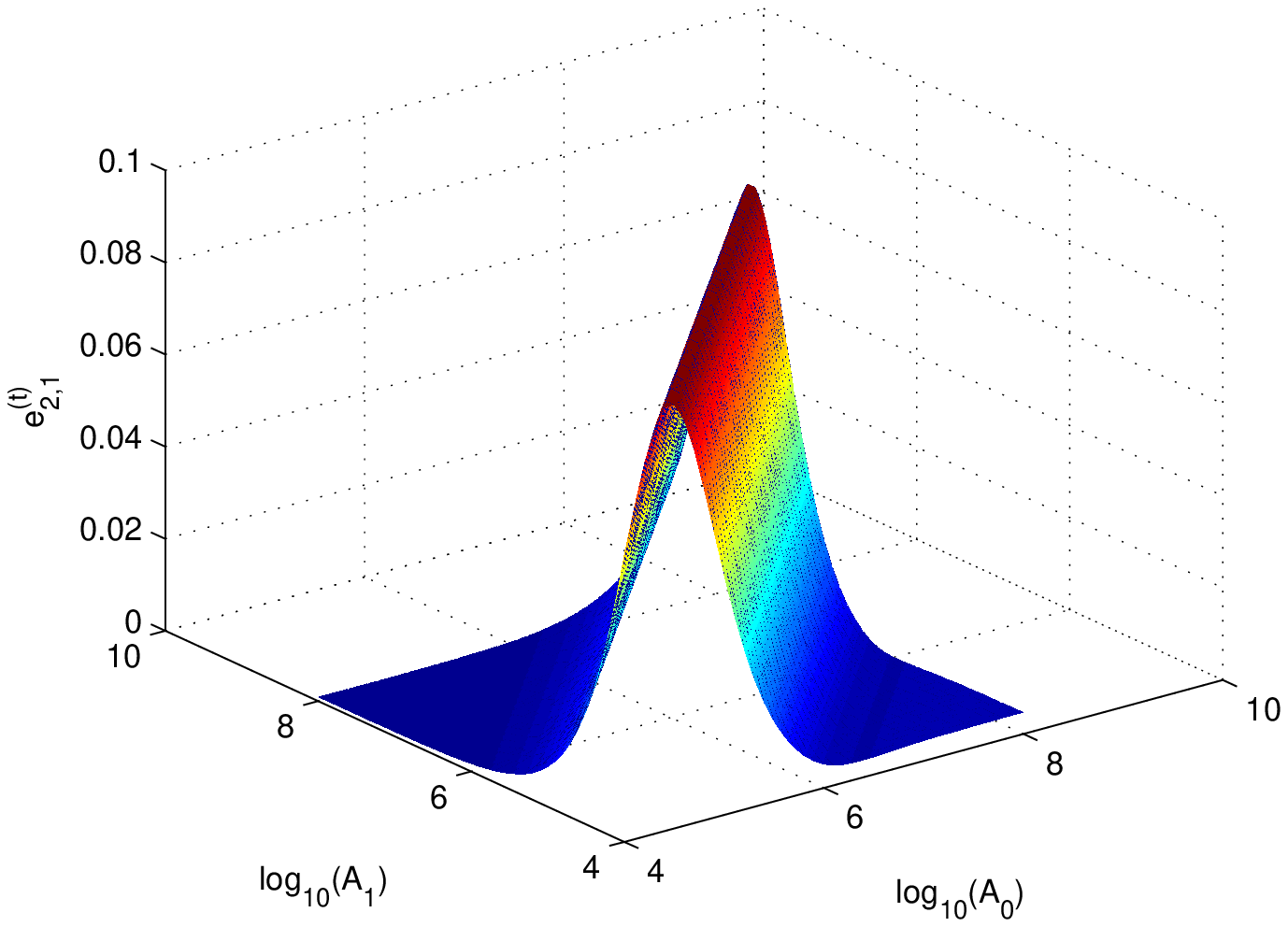}}
          \subfigure[$e^{(t)}_{2,2}$]
             {\includegraphics[width=0.40\textwidth]{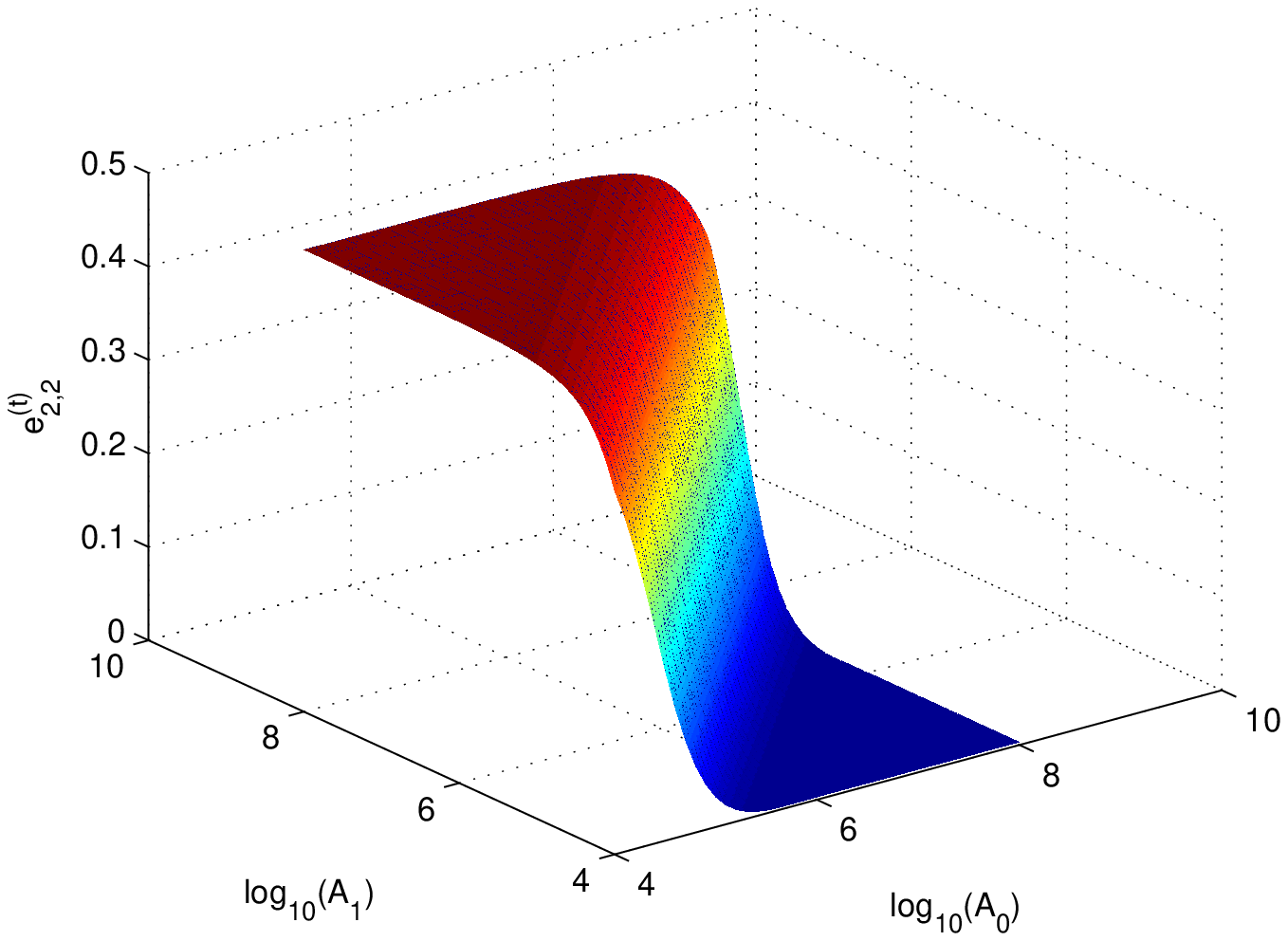}}
       \caption{The efficiency of the transmitted propagating waves at the frequency $2\omega_I$. Schematically, when $A_0>A_1$ (resp. $A_0\sim A_1$, $A_0<A_1$), the $0^{th}$ (resp. $1^{st}$, $2^{nd}$) order is favored. That $e^{(t)}_{2,1}$ is peaked around $A_0=A_1$ was a surprise for us.}
       \label{figSlab2t}
\end{figure}

\begin{figure}
\centering
          \subfigure[$e^{(t)}_{3,0}$]
             {\includegraphics[width=0.40\textwidth]{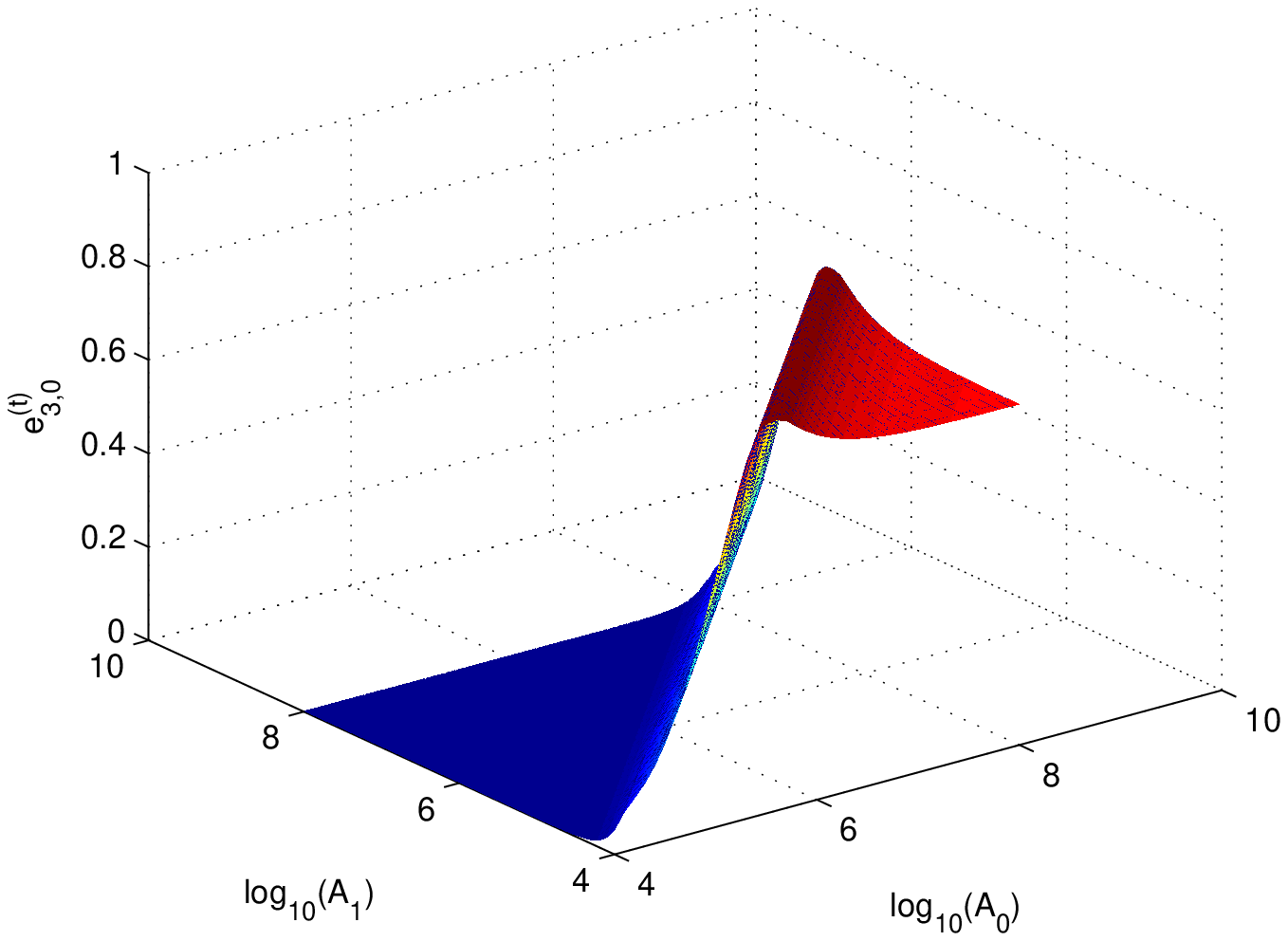}}
          \subfigure[$e^{(t)}_{3,1}$]
             {\includegraphics[width=0.40\textwidth]{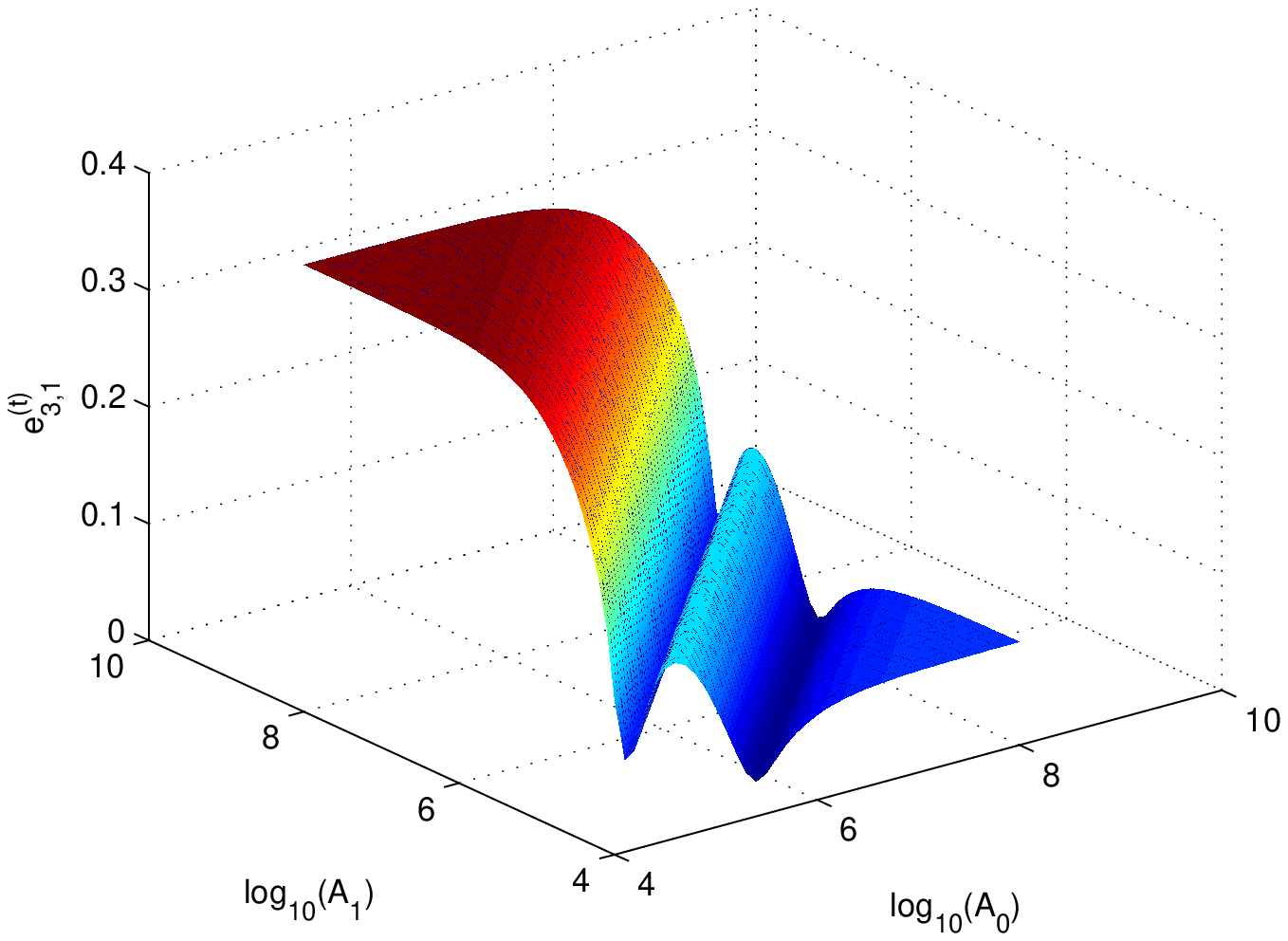}}
          \subfigure[$e^{(t)}_{3,2}$]
             {\includegraphics[width=0.40\textwidth]{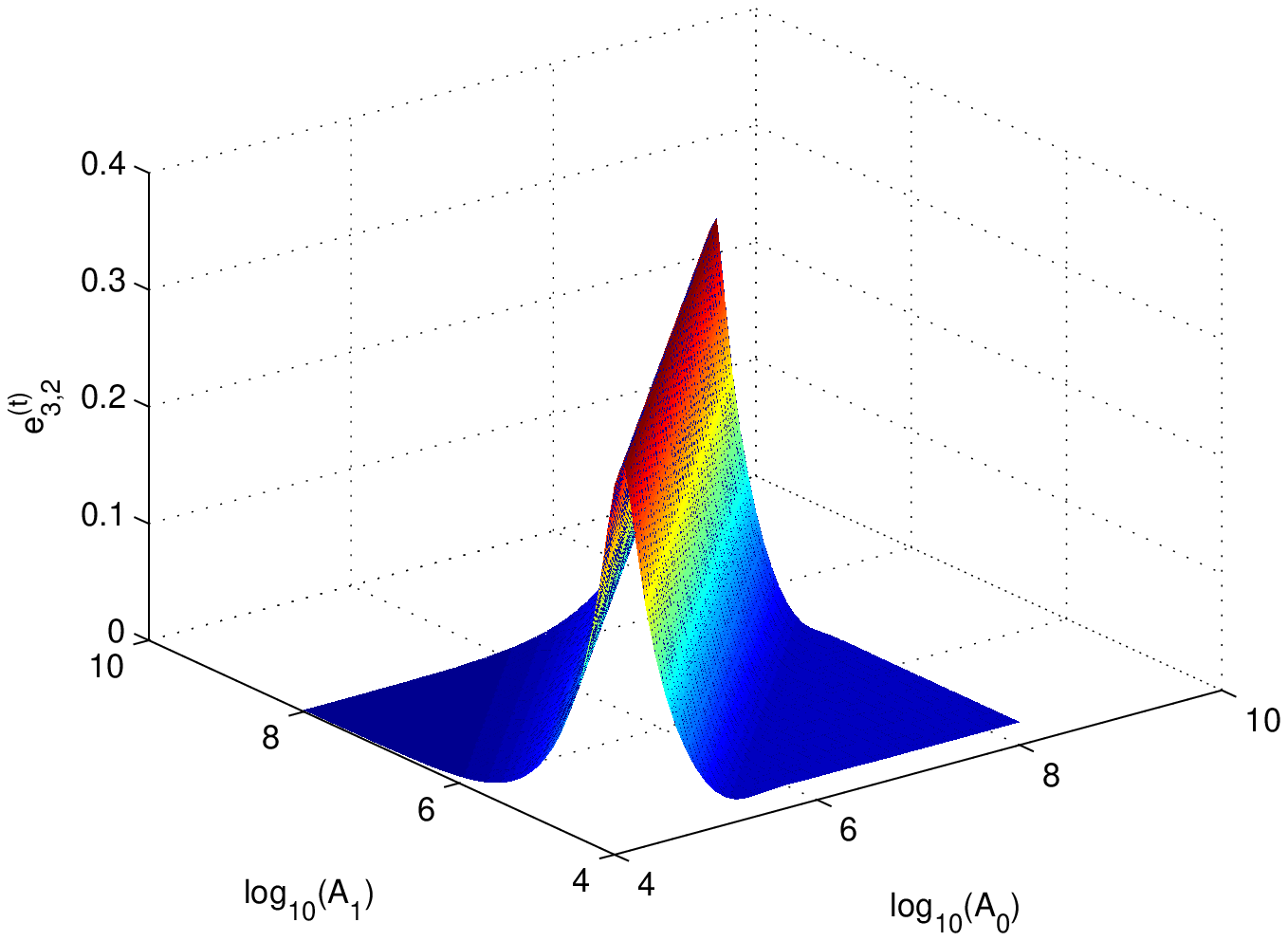}}
          \subfigure[$e^{(t)}_{3,3}$]
             {\includegraphics[width=0.40\textwidth]{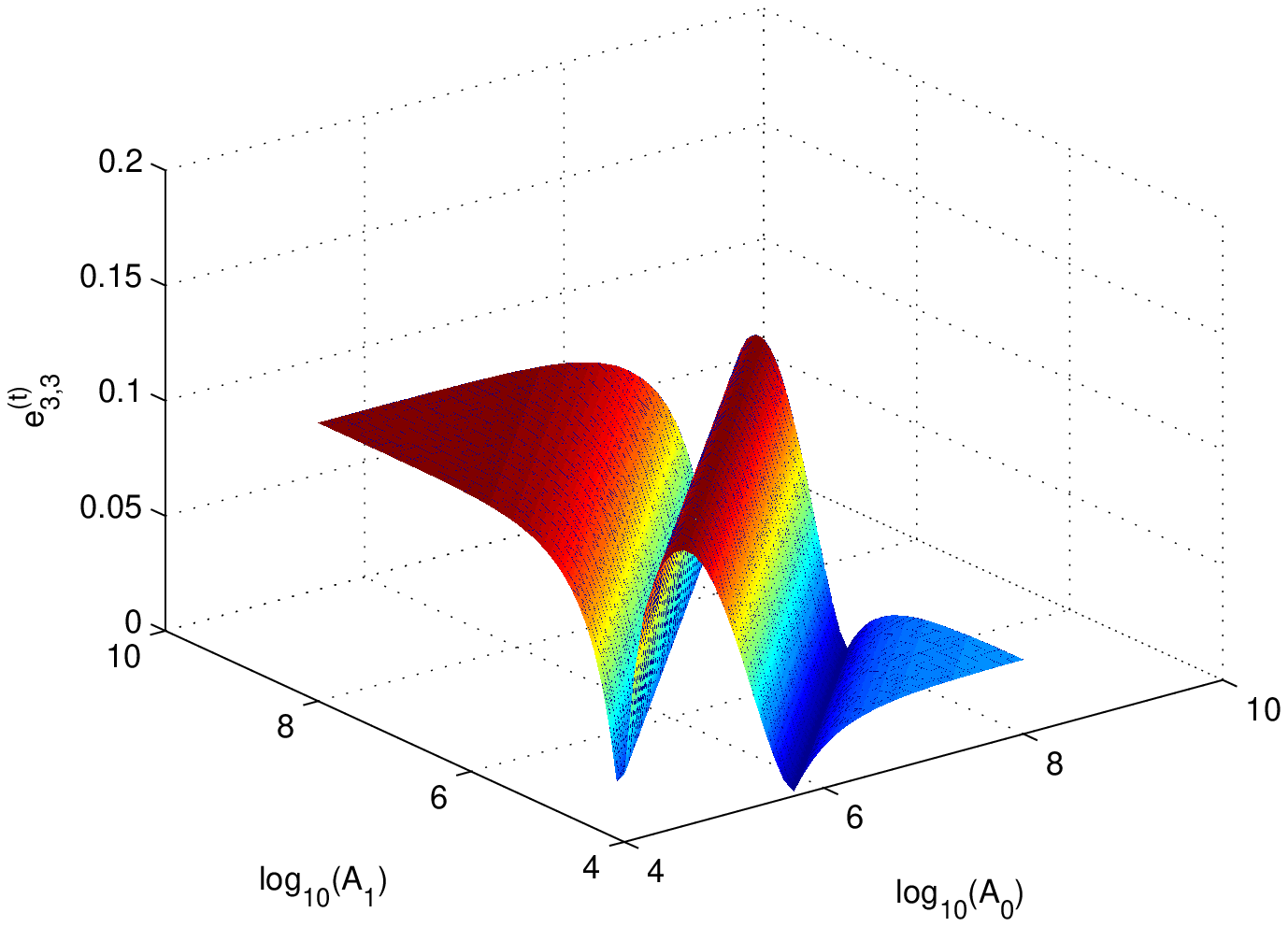}}
       \caption{The efficiency of the transmitted propagating waves at the frequency $3\omega_I$. These maps present large resonances (see the scale for $e^{(t)}_{3,2}$).}
       \label{figSlab3t}
\end{figure}

The second thing we note is that the amplitude of the $n$-th order
of the $p$-th harmonic (that is $b^{(r)}_{p,n}$ or $b^{(t)}_{p,n}$)
is not monotonic in the amplitude of the incident field, as is seen
in the figures \ref{figSlab1t}-\ref{figSlab3t} (due to the
reflection symmetry along the $y$-axis of the system,
$b^{(t)}_{p,-n}=b^{(t)}_{p,n}$ and $b^{(r)}_{p,-n}=b^{(r)}_{p,n}$).
In the figures are represented the efficiency of the transmitted
waves - the reflected ones show similar behaviors - defined by:

$$e^{(a)}_{p,n}:=\frac{|b^{(a)}_{p,n}|}{\sum_{n\in\mathcal{U}_p}|b^{(a)}_{p,n}|}$$
for $a\in\{r,t\}$. It thus gives the part of the intensity of the
reflected or transmitted wave at pulsation $p\omega_I$ that escapes
along the $n$-th order.

\section{Concluding Remarks}

We present numerical evidences that the scattering of several waves
on a nonlinear slab, whose thickness is of the order of the
wavelength, shows rich phenomena. This system is studied through a
set of equations obtained from a rigorous method, given in a
companion paper. The response of the induced grating is far from
being monotonous in the amplitudes of the pump wave. We now look for
an experimental test.

%
%
%
%
%
%
%
%

%

\vspace{.5cm}%
\paragraph{Aknowledgement} The authors are grateful to S. Brasselet and A. Ferrando for their comments about this manuscript.

\bibliographystyle{unsrt}

\bibliography{biblio}

\end{document}